\documentclass[aps,twocolumn,epsfig,graphics,showpacs,floatfix,mathbbm]{revtex4}

\usepackage{amsmath,amsfonts,amssymb,graphics,graphicx,epsfig,color,times,bbm}

\begin{document}

\bibliographystyle{apsrev}

\newcommand{\proofend}{\hfill\fbox\\ \smallskip  }

\newtheorem{conjecture}{Conjecture}
\newtheorem{corollary}{Corollary}
\newtheorem{theorem}{Theorem}
\newtheorem{proposition}{Proposition}
\newtheorem{definition}{Definition}
\newtheorem{lemma}{Lemma}

\newcommand{\bra}[1]{\langle #1|}
\newcommand{\ket}[1]{|#1\rangle}
\newcommand{\braket}[2]{\langle #1|#2\rangle}
\newcommand{\tr}{\text{tr}}

\newcommand{\nn}{\mathbb{N}}
\newcommand{\rr}{\mathbb{R}}
\newcommand{\cc}{\mathbb{C}}
\newcommand{\id}{\mathbb{I}}

\title{\Large Correlated entanglement distillation \\
and the structure of the set of undistillable states}

\author{\large F.G.S.L.\ Brand\~ao and J.\ Eisert\\
{\it\normalsize Institute for Mathematical Sciences, Imperial College London, London SW7 2PE, UK}\\
{\it\normalsize 
QOLS, Blackett Laboratory, Imperial College London, London SW7 2BW, UK}
}

\date{}

\begin{abstract}
We consider entanglement distillation under the assumption
that the input states are allowed to be correlated among each other. 
We hence replace the usually considered independent and identically-distributed hypothesis by the weaker assumption of merely 
having identical reductions. We find that whether a state 
is then distillable or not is only a property of these reductions,
and not of the correlations that are present in the input state.
This is shown by establishing an appealing relation between the 
set of copy-correlated undistillable states and the standard set of 
undistillable states: The  former turns out to be the convex hull of the latter.
As an example of the usefulness of our approach to the study of entanglement 
distillation, we prove a new activation result, which generalizes earlier findings: 
it is shown that for every entangled state $\sigma$ and  every $k$, there exists a 
copy-correlated $k$-undistillable state $\rho$
such that $\sigma \otimes \rho$ is single-copy distillable. 
Finally, the relation of our results to the conjecture about the 
existence of bound entangled states with a non-positive 
partial transpose is discussed. 
\end{abstract}

\maketitle

\section{Introduction}

The concept of entanglement is at the root of 
the field of quantum information science.
Entanglement is thought to render the envisioned
quantum computer more powerful than its classical counterpart, and
has  in some sense to be present to make sure that one can distill
a secure classical key in quantum key distribution. Yet, 
entangled quantum states are 
not defined via their immediate usefulness for 
quantum information purposes, but rather 
via the way they are prepared:
a quantum state is called {\it entangled} if it is not merely
{\it classically correlated}, so if it cannot
be prepared  -- in a 
in a distributed laboratories paradigm -- with local 
quantum operations alone, making use of 
classical shared randomness \cite{Werner,HorodeckiRMP}. These
classically correlated states are hence exactly
those states that can be prepared with
local distributed physical devices. This definition 
in terms of the very preparation procedure, needless
to say, does not imply per se the usefulness of the
entanglement.

In turn, ``useful entanglement'' in a distant laboratories paradigm
may be identified with the concept of {\it distillable
entanglement} \cite{ben}: a quantum state $\rho$
is called distillable if a supply of states which are
independent and identically-distributed -- in other words 
$\rho^{\otimes n}$ -- can be transformed into fewer
almost perfect maximally entangled states,
again using only local quantum operations and classical
communication. Such maximally entangled states of qubit
pairs give immediately rise to a secret bit of key in 
quantum key distribution, or may form the resource in 
quantum state teleportation. A key assumption in such a distillation
process is that the source produces identical uncorrelated 
specimens. 

An interesting generalization of this paradigm is the one in which the several copies of the state $\rho$ are not completely independent
\cite{IID}.
It is worthwhile both from a fundamental and practical point of view to 
study the effects of correlations among 
the copies of $\rho$ on its distillability properties. 
For each natural number $n$, instead of considering 
the usual $n$ uncorrelated copies of the state in question, $\rho^{\otimes n}$, we consider that arbitrary correlations 
exist among those. Hence, the full state is characterized by a density matrix $\omega_n$ with the requirement that $\tr_{\backslash k}(\omega_n) = \rho$ for every $1 \leq k \leq n$, where $\tr_{\backslash k}$ stands for the partial trace of all the copies except the $k$-th. Clearly there are several distinct choices for $\omega_n$, representing the different ways in which the $n$ copies might be correlated. An interesting question in this respect is to classify the set of states for which correlations among its copies can rule out the possibility of obtaining useful entanglement by means of entanglement distillation. It is natural to expect that such correlations could have quite a drastic effect and, hence, that the set of \textit{copy-correlated undistillable states} would be much larger than the usual set of undistillable states. Somewhat 
surprisingly, it turns out that the existence of correlations do not influence to a very large extend whether distillation can be successfully implemented or not. In fact, the state $\omega_n$ does not even
have to be assumed to the permutation-symmetric: Whether the correlated input is distillable
or not merely depends on the reduction $\rho$, and not on the correlations.
We prove that the set of copy-correlated undistillable states is given by the 
convex hull of the set of undistillable states, therefore providing a new 
characterization for the latter. 
 
At the core of this result is, of course, the characterization of the set of undistillable states. One of the key results in entanglement theory is that not every entangled state is distillable, 
demonstrating that there is a kind of {\it bound
entanglement} in nature \cite{hor2}. 
In turn, a certain very simple
criterion was found to be intimately
related to distillability: that of the positivity of 
the partial transposition, obtained by transposition in only
one part of a composite bi-partite system \cite{hor2,Peres}. 
A state that has a positive partial transposition 
is never distillable. What remained a quite
notorious question is whether so-called NPPT 
bound entangled states exist, so states which are not distillable 
but nevertheless exhibit a non-positive partial 
transpose. Its existence has various
ramifications in quantum information science 
\cite{HorodeckiRMP,ShorNPPT,CiracNPPT,ShorCont,NPPTAct,NPPTAct2} -- and would rule
out the appealing feature of being able to test for undistillability
by means of such a simple test as by computing the spectrum of 
the partial transpose.

One direct implication of our result is that if every 
undistillable state is PPT, then arbitrary correlations among 
the copies of the state in question have no effect at all in whether 
distillation can be successfully implemented. We however find that the 
characterization of the set of copy-correlated undistillable 
states we establish actually gives strong indications on the existence of NPPT bound entangled states. 
This is accomplished by proving a new entanglement 
activation result, which generalizes previous findings \cite{Masanes} and 
points towards an activation process involving two undistillable states which 
would in particular imply the 
existence of NPPT states. 

Entanglement activation \cite{ShorCont, NPPTAct, NPPTAct2,Masanes,hor3,Activation, Ishizaka1, Ishizaka2, Masanes2, Brandao1} is the process in which an entangled state, 
which by itself would be useless for a given task e.g. teleportation, can be activated and 
used as an a resource when processed together with a second state. In the first example 
of such a phenomenon \cite{hor3}, it was shown that a certain PPT bound entangled state could be 
employed in order to increase the fidelity of teleportation of a second state. In Ref.\ \cite{Masanes},
in turn, this result was shown to be a general feature of bound entangled states: any entangled state
can increase the fidelity of teleportation of a second state. In this work we generalize such a result, proving that for every $k$, the set of copy-correlated $k$-undistillable states, which are states for which 
correlations among $k$ copies of them can prevent the possibility of obtaining 
a two qubit entangled state by stochastic local operations and classical communication (SLOCC), contains states capable of activating every entangled state. This new activation result we demonstrate then strongly suggests that a similar result might hold true by considering the set of copy-correlated undistillable states itself, which together with its identification with the convex-hull of the set of undistillable states, would be sufficient to prove the existence of NPPT states.  

The structure of this paper is the following. In Section 2 we define the main quantities which we will be concerned with. In Section 3, in turn, we state our main results, which are proved subsequently in Sections 4, 5, and 6. In Section 7 we discuss the connections of our results with the conjecture about the existence 
of NPPT bound entangled states. Finally, we present the summary of this
work together with some further conclusions in Section 8.

\section{Distillability and copy-correlated distillability}

We start by defining the objects we will be using frequently.
${\cal H}=\cc^d\otimes \cc^d$ will denote the Hilbert space
of a bi-partite $d\times d$-dimensional quantum system. 
The state space over ${\cal H}$ is written as
${\cal D}({\cal H})$. Of central interest will be states
$\rho$
on ${\cal H}^{\otimes k}$ for some $k$
that are {\it permutation-symmetric}: This means that 
when permuting any of the $k$ bi-partite quantum states,
the state is left unchanged under the standard representation
of the symmetric group $S_k$ over ${\cal H}^{\otimes k}$,
\begin{equation}
	\rho = P_\pi \rho  P_\pi,
\end{equation}
where $P_\pi$ is the representation in ${\cal H}^{\otimes k}$ of an 
arbitrary element $\pi$ of $S_k$. We define the symmetrization operation as 
\begin{equation}
\hat{S}_k(\rho) := \frac{1}{n!} \sum_{\pi \in S_k} P_\pi \rho  P_\pi.
\end{equation}
The set of 
permutation-symmetric states will be denoted as
${\cal S}_k({\cal H}^{\otimes k})\subset {\cal D}({\cal H}^{\otimes k})$.
We will also freely make use of partial traces over part of 
systems: $\tr_{\backslash 1}$, for example, will refer to the 
partial trace over all but the first $d\times d$-dimensional
bi-partite quantum system. 

Note that an early example
of entanglement distillation of permutation-symmetric states
has been considered in Ref.\ \cite{Old}. There, permutation symmetry
was even considered when permuting each of the $k$ 
subsystems individually, 
\begin{equation}%
	\rho' := \frac{1}{n!^2}\sum_{\pi,\pi'\in S_k}%
	(P_\pi\otimes P_{\pi'})\rho (P_\pi\otimes P_{\pi'}),
\end{equation}
which induces an even higher degree
of symmetry than randomly permuting the $k$ bi-partite
systems, but is included in the above case.

The distillability problem can be cast in terms of 
a notion that renders it more accessible using the
techniques presented later in this work: it is related to the 
so-called {\it SLOCC singlet fraction}, 
\begin{equation}\label{SF}%
	F_{2}(\rho) := \sup_{A, B} \frac{\tr[(A  %
	\otimes B)^{\cal y}\rho (A \otimes B) \phi_{2}]}
	{\tr[(A \otimes B)^{\cal y}\rho (A \otimes B) ]},
\end{equation}
where $A$ and $B$ act on local parts of a bi-partite
system with Hilbert space ${\cal H}$ and $\phi_2 := 
\sum_{i, j=0}^1 
\ket{i, i}\bra{j, j}/2$ is the projector onto the two qubit maximally 
entangled state. In this language,
a state $\rho$ is distillable if and only if 
$F_{2}(\rho^{\otimes n}) > 1/2$ for some $n\in \nn$. 
In turn,  a state $\rho$ is called $n$-undistillable if 
$F_{2}(\rho^{\otimes n}) = 1/2$. Finally, the set of undistillable states is composed of all the states that are $n$-undistillable for all $n \in \mathbb{N}$. We denote the set of $n$-undistillable states by ${\cal C}_n$ and the set of undistillable states by ${\cal C}$.

The central object of this work is the generalization of such sets to the case where correlations among the several copies of the state might be present. We are interested in the worst case scenario and say that a state $\rho$ is copy-correlated $k$-undistillable if there is a $1$-undistillable state $\omega_k \in {\cal D}({\cal H}^{\otimes k})$ such that $\tr_{\backslash m}(\omega_k) = \rho$ for every $1 \leq m \leq k$. In other words, if we can add correlations to the $k$ copy state $\rho^{\otimes k}$, forming the state $\omega_k$, such that no two qubit entanglement can be extracted from $\omega_k$, we say that $\rho$ is copy-correlated $k$-copy undistillabe. It is clear that if such an extension exists, then $\hat{S}_k(\omega_k)$ is also a valid 
extension. It hence follows that w.l.o.g. we can define the set of copy-correlated $k$-undistillable states as 
\begin{definition}[Copy-correlated $k$-undistillable]
We say that a bi-partite state $\rho \in {\cal D}({\cal H})$ is copy-correlated $k$-undistillable if it has a permutation-symmetric extension $\omega_k \in {\cal D}({\cal H}^{\otimes k})$ which is single-copy undistillable. We denote the set of all such states by ${\cal T}_k$, i.e.
\begin{eqnarray}
	{\cal T}_k &:=& \{ \rho \in {\cal D}({\cal H}) 
	: \exists \hspace{0.1 cm} \omega_k \in 
	{\cal S}_k({\cal H}^{\otimes k}) 
	\cap {\cal C}_1({\cal H}^{\otimes k}) \hspace{0.1 cm} 
	\nonumber\\	
	&&\text{s.t.} \hspace{0.2 cm} 
	\rho = \tr_{\backslash 1}(\omega_k) \}.
\end{eqnarray}
\end{definition}

In the same way as one defines undistillability
as $k$-undistillability for all $k$, one can introduce an analogous definition in the 
copy-correlated case 

In more physical terms, the setting of copy-correlated distillation 
is the following: One considers sequences of sources, each
producing permutation-symmetric
correlated bipartite states entailing $k$ pairs
each. This is the natural setting when the source
produces entangled pairs at once, but the 
physical process achieving 
this leads to not entirely uncorrelated specimens. 
Still, for the reductions to be identical 
and equal to some $\rho$ is still 
a reasonable assumption (and the state can also be 
twirled over the symmetric group to make the reductions
identical). In Ref.\ \cite{Datta},
this concept of formation and distillation beyond
i.i.d.\ sources has also been discussed in the pure-state
case. Note also that the correlations between 
the copies can be arbitrarily strong (except that due to 
monogamy constraints, the resulting state will eventually
become copy-correlated undistillable). We do not impose
any restrictions to the kind of correlations allowed. If there
are no correlations, the usual concept of distillation is
recovered.

The parties doing the distillation based on such
a source will for a finite $k$ clearly not be able to do
a quantum state tomography to find out $\rho$: They 
will simply be promised the source to have that property.
This is the natural setting of discussing entanglement distillation
in the presence of cross-copy correlations and memory
effects. Note that we will not be interested in distillation rates
in this work, but just in distillability as such. This naturally
links to the concept of undistillability:

\begin{definition}[Copy-correlated undistillable]
A state $\rho \in {\cal D}({\cal H})$ is said to be copy-correlated undistillable if it is copy-correlated $k$-undistillable for every $k \in \mathbb{N}$. We denote the set of copy-correlated undistillable states by ${\cal T}$, i.e.
\begin{equation}
{\cal T} := \bigcap_{k \in \mathbb{N}^{*}} {\cal T}_k.
\end{equation}
\end{definition}

In words, a state $\rho$ belongs to ${\cal T}$ if for every number of copies of 
the state one can add correlations among them so that no useful entanglement 
can be establish at all. 

This approach
seems interesting for two reasons: 
One the one hand, this is a natural setting to consider, as the 
assumption of having entirely uncorrelated specimen 
at hand in entanglement distillation may be an unacceptably
restrictive one. On the other hand, as we will see, we can use
this concept as a novel mathematical tool to grasp the structure of the set of 
undistillable
states.

Equipped with these definitions, are are now in the position to
state our main results and present the proofs.

\begin{figure}
 \centerline{\includegraphics[width=6.5cm]{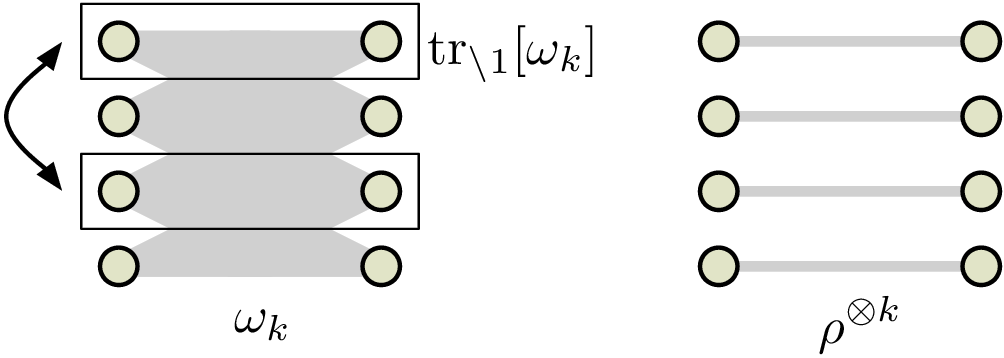}}
  \caption{This figure represents the setting of copy-correlated
  $k$-undistillable states. The reduction to the first bi-partite
  system is $\tr_{\backslash 1}[\omega_k]$, the state is
  invariant under permutations of the bi-partite systems. To the
  right the i.i.d case $\rho^{\otimes k}$.}
\end{figure}

\section{Main results}

The first result concerns the relationship
between copy-correlated undistillable states and the
undistillable states in the ordinary i.i.d. sense:
We find that the set of copy-correlated undistillable states is
nothing but the convex hull of 
the set of undistillable states. Since the latter 
set is possibly non-convex
(a property related to the existence of NPPT bound 
entanglement \cite{ShorCont}), the convex hull of this set might, however, 
indeed be different of the set itself.

\begin{theorem}[Undistillable and copy-correlated 
undistillable states]\label{Relate}
The set of copy-correlated undistillable states is equal to the convex-hull 
of the set of undistillable states:
\begin{equation} 
{\cal T} = co({\cal C}).
\end{equation}
\end{theorem}

Our proofs will make repeated
use of convex analysis \cite{Convex}, 
and extend ideas of employing
convex cones of Ref.\ \cite{Masanes,Masanes2} 
to the asymptotic
setting. The {\it dual cone} of the set ${\cal C}$, for example,
is defined as
\begin{equation}
	{\cal C}^\ast := \left\{
	X\geq 0: \tr[X \rho]\geq 0\, \forall\, \rho\in {\cal C} 
	\right\}.
\end{equation}
Theorem \ref{Relate} then has the following
immediate consequence.

\begin{corollary}[Characterization of the set of undistillable states]
The dual cone of the set of undistillable states can be characterized as follows
\begin{equation}
{\cal C}^* = \overline{\bigcup_{k \in \mathbb{N}^*} {\cal T}_k^*}.
\end{equation}
\end{corollary}

In other words, we have fully characterized the dual cone of the
set of undistillable states in terms of sets which are easily specified. 

In the above framework, a standard maximally entangled state
of dimension $\cc^2\otimes \cc^2$ is taken, as the {\it singlet fraction}
is taken as the figure of merit. Note that we aim for the question
of obtaining such a singlet in a distillation protocol, but do not
study rates of distillation here. We emphasize, however, that
once $\rho\not\in {\cal T}$, one can distill an arbitrary good approximation of a maximally
entangled output of arbitrary dimension. This is the 
content of the next Corollary.

\begin{corollary}[Distillation with arbitrary output dimension]
Let $\rho\in {\cal D}({\cal H})$ be a state for which 
$\rho\not\in {\cal T}$. Then, for every sequence of states $\{ \omega_n \}$ with 
reductions equal to $\rho$, every integer $D$, and every $\lambda \in [1/D, 1)$, there 
is an integer $n$ and an SLOCC map 
such that   
\begin{equation}
F_D(\omega_n) > \lambda.
\end{equation}
\end{corollary}

The second main result is a generalization of the activation result 
proved in Ref.\ \cite{Masanes}. It indicates the power of 
copy-correlated $k$-undistillable states to serve as activators
to make states distillable. In fact, this is true on the single-shot
level, so the resulting states are even single-copy distillable
\cite{Comm}.
Again, 
the interesting aspect of this result is that it is a statement
on asymptotic entanglement manipulation. But the whole
asymptotic aspect is hidden in the characterization of
the set of copy-correlated undistillable states: As an activation
result, it refers to an operation on a single specimen alone.

\begin{theorem}[Main result on activation of entanglement]
For every entangled state $\rho \in {\cal D}({\cal H})$ and every $k \in \mathbb{N}^*$ there is a copy-correlated $k$-undistillable state $\sigma$ such that the joint state $\rho \otimes \sigma$ is single-copy distillable, i.e.
\begin{equation}  
F_{2}(\rho \otimes \sigma) > \frac{1}{2}.
\end{equation}
\end{theorem}

As ${\cal C}_1 = {\cal T}_1$, the main result of Ref.\ \cite{Masanes} is a particular case of Theorem 2. There is an immediate 
Corollary of the previous result which we can state as follows.

\begin{corollary}[Activation using convex combinations]
For every entangled state $\rho \in {\cal D}({\cal H})$ and any 
$\varepsilon > 0$ there is a single-copy undistillable state $\sigma$ 
such that
\begin{enumerate}
	\item We can find a probability distribution 
	$\{ p_i \}$ and a set of undistillable states $\{ \rho_i \}$ satisfying
\begin{equation}	
	\bigl \Vert \sigma - \sum_i p_i \rho_i	\bigr \Vert_1 \leq \varepsilon,
\end{equation}	
  \item The joint state $\rho \otimes \sigma$ is single-copy distillable.
\end{enumerate}
\end{corollary}

This Corollary is a direct consequence of Theorem 2 
and a standard result of convex analysis stating that a family of 
closed convex sets $\{ A_i \}$ such that $A_{i + 1} \subseteq A_i$ 
converges to their intersection with respect to the Hausdorff 
distance \cite{Kuratowski}.

\begin{figure}
 \centerline{\includegraphics[width=6.5cm]{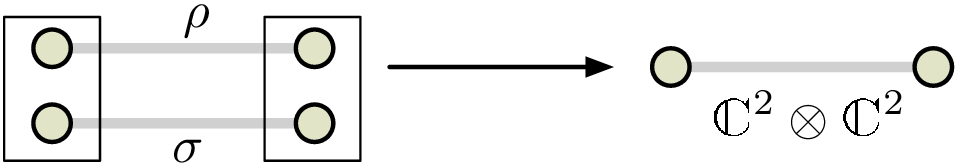}}
  \caption{Activation of entanglement: For any entangled
  state $\rho$ one can find a copy-correlated $k$-undistillable 
  activator $\sigma$ such that the joint state is single-copy
  distillable. This holds true for any $k$. }
\end{figure}

Motivated by these findings, we are -- once again -- led
to the following conjecture \footnote{Note that recent
work on the existence of NPPT bound entangled states
has introduced interesting ideas \cite{Simon}, 
but the argument
as such is not correct in its
conclusion. To assume that the 
extreme points of the 
convex sets constructed in Ref.\ \cite{Simon}
are pure is a restriction of generality. For a discussion,
see also Ref.\ \cite{HorodeckiRMP}.
}.
Note the strong similarity with the previous statement. 

\begin{conjecture}[Existence of NPPT bound entanglement]
For every entangled state $\rho \in {\cal D}({\cal H})$ there is an undistillable 
state $\sigma$ such that the joint state $\rho \otimes \sigma$ is 
single-copy distillable, i.e.
\begin{equation}  
F_{2}(\rho \otimes \sigma) > \frac{1}{2}.
\end{equation}
\end{conjecture}

This statement would clearly indicate the existence of NPPT
bound entangled states. 
To see this, let us assume that the contrary is true, 
that all undistillable states have a positive
partial transpose. Yet, according to the above
conjecture, for any PPT bound entangled state $\sigma$
there exists an undistillable state $\rho$  -- hence also a PPT state
--  for which $F_{2}(\rho \otimes \sigma) > 1/2$. This leads to a 
contradiction, as $\rho\otimes \sigma$ has in turn a positive
partial transpose, which implies that 
\begin{equation}	
	F_{2}(\rho \otimes \sigma) 
	= \frac{1}{2},
\end{equation}	
in contradiction to the assumption. Hence, our
above result is also aimed at providing a new instrument in
tackling the old conjecture on the existence of NPPT
bound entanglement \footnote{For recent
progress on the question of the existence
of NPPT distillable entanglement, see also 
Ref.\ \cite{Problem}.}.

\section{Proof of Theorem 1}

We will now proceed by proving the validity of the two theorems.
We start with the preparation of the proof of Theorem 1. This 
argument will make use of de-Finetti and
large deviation techniques \cite{Fuchs,Christandl,Renner}. 
The first statement that is of
use is borrowed from Ref.\ \cite{Christandl}. Note that here
it is stated with respect to bi-partite systems, which is
responsible for the obvious difference in the scaling in 
the dimension $d$.

\begin{theorem} 
[Quantum finite de Finetti theorem \cite{Christandl}] \label{definetti} 
Let $\omega_n$  
be a permutation-symmetric state $\omega_n\in 
{\cal S}_n({\cal H}^{\otimes n})$ and let 
$k \leq n$. Then there exists a 
probability distribution $P$ over state space ${\cal D}({\cal H})$
such that 
\begin{equation}
\Vert \tr_{k + 1, \dots, n}(\omega_n) - \int P(\rho) 
\rho^{\otimes k} d\rho \Vert_1 \leq \frac{4 d^4 k}{n}.
\end{equation}
\end{theorem}

With the help of the previous statement, we can characterize
the set of copy-correlated undistillable states. Note that the
following lemma does not constitute an assumption on the
specific form of the correlations between the copies produced
by the source, but it is a result that holds true as a consequence
to any input states having such correlations. 

\begin{lemma}[Set of copy-correlated undistillable states] \label{finetti}
A  state $\sigma\in {\cal D}({\cal H})$ 
belongs to ${\cal T}$ if and only if there 
exists a probability distribution $P$
over state space ${\cal D}({\cal H})$  such that
\begin{equation}
\sigma = \int P(\rho)\rho d\rho,
\end{equation}
and
\begin{equation} \label{pin}
\pi_{k} := \int P(\rho) \rho^{\otimes k} d\rho \hspace{0.1 cm} \in \hspace{0.1 cm} {\cal C}_1({\cal H}^{\otimes k}) 
\end{equation}
for every $k \in \mathbb{N}^*$.
\end{lemma}

\textit{Proof:} 
Let $\sigma \in {\cal T}$, then, for each $k \in \mathbb{N}^*$, 
there exists a state
$\omega_k \in {\cal C}_1({\cal H}^{\otimes k})$ such that $\tr_{\backslash 1}(\omega_k) = \sigma$. This is a direct
consequence of the definition of ${\cal T}$.
From Theorem \ref{definetti} it follows that for each $k \geq 1$, 
there exists 
a probability distribution $P_k(\rho)$ such that
\begin{equation}
\Vert \tr_{k + 1, \dots, k^2}(\omega_{k^2}) - \int P_k(\rho) \rho^{\otimes k} d\rho  \Vert_1 \leq \frac{4 d^4}{k}.
\end{equation} 
Let us define
\begin{equation}
\pi^k_{j} := \tr_{j + 1, \dots, k^2} (\omega_{k^2}).
\end{equation}
From the property that the trace norm is contractive under 
completely positive maps, and hence
under partial tracing, we have that for each $j \leq k$,
\begin{equation} \label{EEE}
\Vert \pi^k_j - \int P_k(\rho) \rho^{\otimes j} d\rho  \Vert_1 \leq 
\frac{4 d^4}{k}.
\end{equation} 
Moreover, as locally discarding 
some part of a state amounts to a LOCC operation, 
we have that each 
\begin{equation}
	\pi^k_j\in{\cal C}_1({\cal H}^{\otimes j}). 
\end{equation}
The set of probabilities on the state space is compact in the weak* topology. So there is a probability measure $P$ and a net $k(\alpha)$ of integers such that $k(\alpha) \rightarrow \infty$ and $P_{k(\alpha)} \rightarrow P$. The map 
\begin{equation}
	P\rightarrow \int P(\rho) \rho^{\otimes j} d\rho 
\end{equation}
is continuous, so it also follows that
\begin{equation}
\int P_{k(\alpha)}(\rho) \rho^{\otimes j} d\rho \rightarrow \int P(\rho) \rho^{\otimes j} d\rho 
\end{equation}
Then, by Eq.\ (\ref{EEE}) we find that for every $j \in \mathbb{N}^*$,
\begin{equation}
\pi^{k(\alpha)}_j \rightarrow \int P(\rho) \rho^{\otimes j} d\rho.
\end{equation}
As for every $k(\alpha)$ and $j$, 
\begin{equation}
\tr_{\backslash 1}(\pi^{k(\alpha)}_j) = \sigma \hspace{0.3 cm} \text{and} \hspace{0.3 cm} \pi^{k(\alpha)}_j \in {\cal C}_1, 
\end{equation}
it follows that 
\begin{equation}
\tr_{\backslash 1}\left( \int P(\rho) \rho^{\otimes j} d\rho \right) = \sigma  \hspace{0.3 cm} \text{and} \hspace{0.3 cm} \int P(\rho) \rho^{\otimes j} d\rho  \in {\cal C}_1
\end{equation}
hold true for every $j \in \mathbb{N}^*$. The converse direction of the proof follows directly
from the definition of ${\cal T}$.
\proofend

The next concept that we need is that of a minimal informationally complete POVM. An 
{\it informationally complete POVM} 
in ${\cal B}(\cc^{m})$ is defined as a set of positive semi-definite operators $A_i$ forming a resolution of the identity, i.e., 
satisfying
\begin{equation}
	\sum_i A_i = \id.
\end{equation}
In addition, $\{ A_i \}$ has to form 
a basis for ${\cal B}(\cc^{m})$. An 
informationally-complete POVM is said to be {\it minimal}, 
in turn, when each 
operator $X \in {\cal B}(\cc^{m})$ 
is uniquely determined by the expectation values 
$\tr[A_i X]$. We will make use of a construction of minimal informationally 
complete POVMs presented in Ref.\  
\cite{Renner}, valid for all dimensions $m$. 

We say that a family $\{ A_i \}$  of elements from 
${\cal B}(\cc^m)$ is a {\it dual} of the a 
family $\{ A_i^* \}$ if for all $X \in {\cal B}(\cc^m)$,
\begin{equation} \label{dualbasis}
X = \sum_{i} \tr[A_i X] A_i^*.
\end{equation}
The above equation implies in particular that the operator 
$X$ is fully determined by the expectations values $\tr[A_i X]$. 
Finally, if $\{ A_i \}$ and $\{ B_j \}$ are informationally complete POVMs on ${\cal B}(\cc^m)$ and 
${\cal B}(\cc^l)$, then $\{ M_{i,j} \}$, defined by 
\begin{equation}\label{Prod}
	M_{i,j} := A_i \otimes B_j, 
\end{equation}
is an informationally complete POVM on ${\cal B}(\cc^m\otimes \cc^l)$.
Before now turning to the proof of Theorem 1, there is
one last ingredient that we need for our argument: It may
be viewed as a variant of a Chernoff bound. Note that this
is a statement on classical probability distributions, not
on quantum states.
 
\begin{lemma} [Variant of Chernoff's bound \cite{Tomas}] \label{renner}
Let $P_X$ be a probability distribution on ${\cal X}$ and let $x$ be chosen according to the $n$-fold product distribution $(P_X)^n$. Then, for any $\delta > 0$, 
\begin{equation} 
\text{Pr}_{x} [|| \lambda_x - P_X  ||_1 > \delta] \leq 2^{- n (\frac{\delta^2}{2 \ln 2} - |{\cal X}|\frac{\log(n + 1)}{n})}.
\end{equation}
Here, $||.||_1$ is the trace distance of two probability distributions and $|{\cal X}|$ is the cardinality of ${\cal X}$. 
\end{lemma}

\textit{Proof of Theorem 1:}
We proceed by showing that 
both $\text{co}({\cal C}) \subseteq {\cal T}$ and ${\cal T} \subseteq \text{co}({\cal C})$ hold true. We start with the first inclusion,
which is quite straightforward. Let $\rho\in {\cal C}({\cal H})$ 
be an undistillable state. By symmetry, it is clearly
true that the state $\rho^{\otimes n}$ belongs to 
${\cal S}_{n}({\cal H}^{\otimes n})$ for all $n$. 
Moreover, $\rho^{\otimes n}$ is by definition not single-copy distillable.
%
Therefore, $\rho^{\otimes n}$ belongs to ${\cal C}_1({\cal H}^{\otimes n})$. 
Hence, for all $n$, 
\begin{equation}
	\rho^{\otimes n} \in {\cal S}_{n}({\cal H}^{\otimes n}) 
	\cap {\cal C}_1({\cal H}^{\otimes n}), 
\end{equation}
from which it follows that $\rho \in {\cal T}$. 
As ${\cal T}$ is a closed convex set, one 
finds that indeed $\text{co}({\cal C}) \subseteq {\cal T}$. 

Let us now
consider the converse inclusion. To this aim, 
let $\pi \in {\cal T}$. Then for each $n \in \mathbb{N}^*$ there exists
a $\pi_n$ given by Eq.\ (\ref{pin}) such that 
\begin{equation}
	\tr_{\backslash 1}[\pi_n] = \pi. 
\end{equation}
Also, Lemma \ref{finetti} defines a probability distribution
$P$ for $\pi$, independent of $n$. Similarly,
for any $n,m\in \mathbb{N}^*$ we find a 
$\pi_{n+m}$.

We will now show that this probability distribution $P$ 
is up to a set of measure zero 
supported only on undistillable states. 
We do this proving that for every 
$n \in \mathbb{N}^*$, the probability function $P(\rho)$ 
vanishes for all 
$n$-distillable states, except from a set of measure zero. 
The ideas of the argument is as follows: We consider
$\pi_{n+m}$, and construct a SLOCC that performs
measurements based on an informationally complete
POVM in the last $m$ systems. Based on this information,
one performs a further operation on the first $n$ systems 
depending whether it is distilable or not.

More specifically,  
for any $n,m\in\mathbb{N}^* $ we define
the SLOCC map 
$\Lambda_{m, n} : 
{\cal B}({\cal H}^{\otimes (m + n)}) \rightarrow 
{\cal B}(\cc^2 \otimes \cc^2)$ as follows: 

\begin{itemize}

\item We first measure the 
informationally-complete POVM $\{ M_{i,j} \}=:
\{ M_{k} \}$ of Eq.\ (\ref{Prod})
individually on each of 
the last $m$ bi-partite systems, 
where $k$ is the joint index labeling the outcomes. 
This is clearly an operation that can be
implemented by means of LOCC: One has to perform
the local POVM on each side. In this way, one can estimate
an empirical probability distribution $P_m(k)$ 
from the relative frequency of the outcomes 
$k$ of the POVM. 
\item 
Then, using Eq.\ (\ref{dualbasis}), we form the operator
\begin{equation}
X_m = \sum_k P_m(k) M_k^* \in {\cal B}({\cal H}).
\end{equation}
Of course, this might not be a valid density operator. 

\item Thus, we define $\sigma_m\in {\cal D}({\cal H})$ 
as the state 
which is closest in trace norm to $X_m$, so as the 
state that minimizes
\begin{equation}
	\left\{
	\|\sigma_m- X_m\|_1:
	\sigma_m \in {\cal D}({\cal H})
	\right\}.
\end{equation}
This is done based on the measurement outcomes
obtained above. If $\sigma_m$ defined in this way is 
not unique, we select one from the respective set of 
solutions. The state
$\sigma_m$ can now either be $n$-distillable or
$n$-undistillable. Note that so far, the only physical
operation performed was the measurement in the 
last $m$ systems. 

\item In the first case, so if
$\sigma_m\in {\cal D}({\cal H})$ is $n$-distillable,
we apply the trace preserving LOCC map 
$\Omega$ on the remaining $n$ systems which 
minimizes the following expectation value:
\begin{equation} \label{Eqq22}
\tr[\Omega(\sigma_m^{\otimes n})(\id/2 - \phi_2)].
\end{equation}
This is the optimal distillation procedure on $n$
copies, $\Omega: {\cal B}({\cal H}^{\otimes n})
\rightarrow {\cal B}(\cc^2\otimes \cc^2)$. 
The map $\rho\mapsto \tr[\Omega(\rho^{\otimes n})(\id/2 - \phi_2)]$,
where $\Omega$ is the trace-preserving LOCC that minimizes 
$\tr[\Omega(\rho^{\otimes n})(\id/2 - \phi_2)]$, is trace-norm
continuous.

\item
In the second case, so
if $\sigma_m\in {\cal D}({\cal H})$ 
is $n$-undistillable, we discard 
the state and replace it by the zero operator on
${\cal H}^{\otimes n}$.
\end{itemize}
This procedure defines our family of SLOCC
operations $\Lambda_{m, n} : 
{\cal B}({\cal H}^{\otimes (m + n)}) \rightarrow 
{\cal B}(\cc^2 \otimes \cc^2)$. 

We know that 
\begin{equation} \label{eq52}
\tr[\Lambda_{n, m}(\pi_{n + m})(\id/2 - \phi_2)] \geq 0,
\end{equation}
for all $m$, as, by Lemma \ref{finetti}, $\pi_{n + m} \in {\cal C}_1({\cal H}^{\otimes{(n + m)}})$.

\begin{figure}
 \centerline{\includegraphics[width=6.5cm]{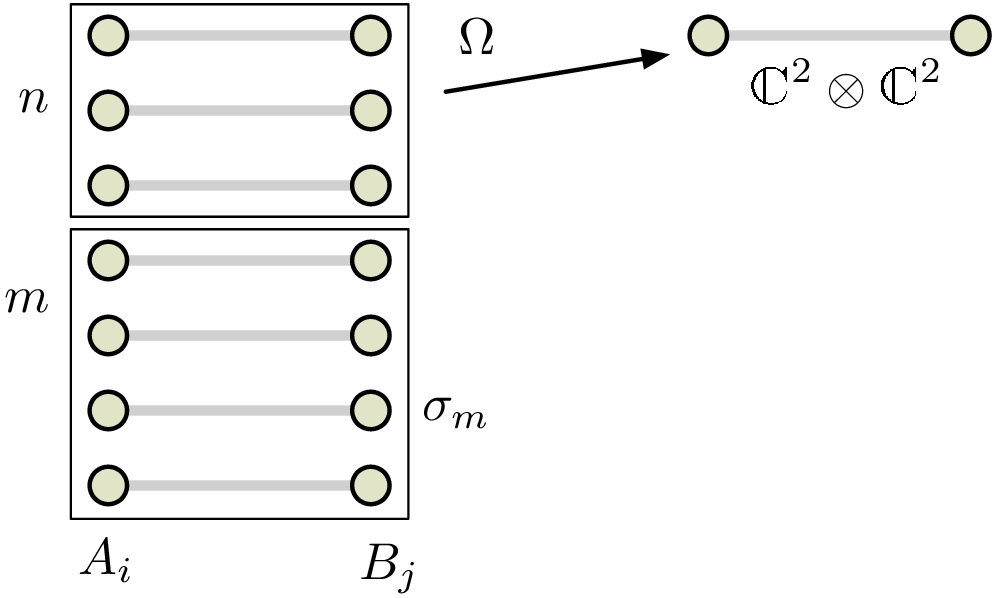}}
  \caption{Procedure followed to define the SLOCC map
  $\Omega$.}
\end{figure}

From Lemma \ref{renner}
we can infer that the probability that the trace norm difference
of the estimated state with the real state is 
larger than $\varepsilon$, for any $\varepsilon > 0$, 
goes to zero when $m$ goes to infinity. So we find 
that the family of functions, defined  
for states $\rho\in {\cal D}({\cal H})$ as
\begin{equation}
f_{m}(\rho) :=  \tr[\Lambda_{m, n}(\rho^{\otimes{(n + m)}})(\id/2 - \phi_2)],
\end{equation}
for fixed $\Lambda_{m, n}$ for any $n,m$, converge pointwise to 
\begin{equation}
f(\rho) := 
\begin{cases}
\tr[\Xi_{\rho}(\rho^{\otimes n})(\id/2 - \phi_2)], & \text{if} \hspace{0.2 cm} \rho \hspace{0.2 cm} \text{is $n$-distillable} \\
0 & \text{otherwise},
\end{cases}
\end{equation}
where $\Xi_{\rho}:{\cal B}({\cal H}^{\otimes n}) \rightarrow 
{\cal B}(\cc^2 \otimes \cc^2)$ is the optimal LOCC map for $\rho^{\otimes n}$, i.e. the LOCC map that 
minimizes 
\begin{equation} \label{Eqq22}
\tr[\Xi(\rho^{\otimes n})(\id/2 - \phi_2)].
\end{equation}
To proceed, we clearly
have the upper bound 
\begin{equation}
	|f_{m}(\rho)| = 
	| \tr[\Lambda_{m, n}(\rho^{\otimes{(n + m)}})(\id/2 - \phi_2)]| 
	\leq 1
\end{equation}	
for every $\rho\in{\cal D}({\cal H})$. This means that the family of functions $\{ f_m \}$ satisfies the requirements
of the Lebesgue dominated convergence Theorem. 
Therefore, we get from Eq.\ (\ref{eq52}) that
\begin{eqnarray} \label{limlebegue}
0 & \leq & \lim_{m \rightarrow \infty} \int P(\rho) \tr[\Lambda_{m, n}(\rho^{\otimes{(n + m)}})(\id/2 - \phi_2)]d\rho \\ &=&  \lim_{m \rightarrow \infty} \int P(\rho) f_m(\rho) d\rho = \int P(\rho) \lim_{m \rightarrow \infty} f_m(\rho) d\rho \nonumber \\ 
&=& \int P(\rho) f(\rho) d\rho \nonumber\\
&=&
\int_{{\cal D}({\cal H})\backslash{\cal C}_{n}({\cal H})} 
P(\rho) \tr[\Xi_{\rho}(\rho^{\otimes n})(\id/2 - \phi_2)] d\rho,\nonumber
\end{eqnarray}
where ${\cal D}({\cal H})\backslash{\cal C}_{n}({\cal H})$ are the
$n$-distillable states.
By definition, we have that for each 
$n$-distillable state $\rho$,
\begin{equation}
	\tr[\Omega(\rho^{\otimes n})(\id/2 - \phi_2)] < 0.
\end{equation}
So we find from Eq.\ (\ref{limlebegue}) that 
$P$ can be non-zero only in a zero measure subset of the set 
of $n$-distillable states. As this is true for an arbitrary $n$, 
we find that $P(\rho)$ must be supported on the set of 
undistillable states.\proofend

\textit{Proof of Corollary 2:} 
We can prove the Corollary by contradiction. Suppose conversely that 
for every $n \in \mathbb{N}$ and every 
SLOCC $\Omega$, $\tr[\Omega(\omega_n)(\lambda\id - \phi_D)] \geq 0$. 
Then we can follow the proof of Theorem 1 to show that
 $\rho \in {\cal T}$, in contradiction with the assumption 
 that it is not. 

The key point is to notice that Theorem 1 also holds if we replace the single copy undistillability condition $F_2(\omega_n) = 1/2$  by 
$F_D(\omega_n) \leq \lambda$, for any integer $D$ 
and $\lambda \in [1/D, 1)$. We only have to modify the fourth 
step of the SLOCC map we defined as follows: we now discard 
the state if the estimated state $\sigma_m$ is such that 
$\tr[\Omega(\sigma_m^{\otimes n})] \leq \lambda$ for every 
SLOCC map $\Omega$, or apply the optimal SLOCC map 
$\Omega$ minimizing $\tr[\Omega(\sigma_m^{\otimes n})
(\lambda \id - \phi_D)]$ otherwise. The proof then proceeds in a completely analogous way.
\proofend

\section{Proof of Theorem 2}

We now turn to the proof of Theorem 2.
We start by proving two auxiliary Lemmas, 
which give a characterization for the elements of the 
dual cones of the sets ${\cal S}_k({\cal H}^{\otimes k})$ 
and ${\cal T}_k({\cal H}^{\otimes k})$, which will
again sometimes
be abbreviated as   ${\cal S}_k$ and
${\cal T}_k$. 

\begin{lemma}[Dual cone of symmetric states] \label{S(Q)=0}                  
If $Q \in ({\cal S}_k)^*$, then                
\begin{equation}      
\hat{S}_{k}(Q) = \frac{1}{n!} \sum_{\pi \in S_k} P_{\pi} Q P_{\pi}\geq 0
\end{equation}                                    
\end{lemma}                 
\textit{Proof:}                        
As $Q \in ({\cal S}_{k})^*$, we have that for every positive semi-definite operator $X \geq 0$ acting on ${\cal H}^{\otimes k}$,                   
\begin{equation}                     
	\tr[ X \hat{S}_{k}(Q)] = \tr[\hat{S}_{k}(X)Q] \geq 0.
\end{equation}                      
This can only be true, however, if
 $\hat{S}_{k}(Q) \geq 0$.
\proofend       

\begin{lemma} [Dual cone of $k$-copy undistillable
states]\label{dual}     
For each $k \in \nn$ and for every element                       
$X$ of $\hspace{0.1 cm}{\cal T}_{k}^{*}$, there exist an 
SLOCC map $\Lambda$ and an operator 
$Q \in ({\cal S}_{k})^*$ such that 
\begin{eqnarray} \label{NF1}                                     
	&& X\otimes \id^{\otimes{( k - 1)}} = \Lambda(\id/2 - \phi_2) + Q.                                       
\end{eqnarray}
\end{lemma}

\textit{Proof:} 
In Ref.\ \cite{lasserre} it has been 
shown that for any two closed convex cones 
$A$ and $B$ defined on a finite dimensional Hilbert space, 
$(A \cap B)^{*} = A^{*} + B^{*}$. It is easily seen that $\text{cone}({\cal S}_k \cap {\cal C}_1) = \text{cone}({\cal S}_k) \cap \text{cone}({\cal C}_1)$,
where
the {\it conic hull} is defined for a set $C$ as
\begin{equation}
	\text{cone}(A) := 
	\biggl\{ \sum_j \lambda_j W_j: 
	\hspace{0.2 cm} \lambda_j \geq 0, \hspace{0.1 cm} 
	W_j \in C \biggr\}.
\end{equation}
Therefore,                                    
\begin{eqnarray}                        
	({\cal S}_{k} \cap {\cal C}_{1})^{*} &=& 
	[\text{cone}({\cal S}_{k} \cap {\cal C}_{1})]^{*} 
	\nonumber\\
	&=& 
	[\text{cone}({\cal S}_{k})\cap \text{cone}({\cal C}_1)]^*
	 = {\cal S}_{k}^* + {\cal C}_{1}^*.
\end{eqnarray}                           
This in turn implies that every element $Y$ of $({\cal S}_{k} \cap {\cal C}_{1})^{*}$ can be written as the right hand side of Eq.\ 
(\ref{NF1}).
We find that if                                        
\begin{equation}                                    
	X \in {\cal T}_{k}^{*}, 
\end{equation}	                                    
then $X \otimes {\id}^{\otimes ( k - 1) }$ is an element of $({\cal S}_{k} \cap {\cal C}_{1})^{*}$. Indeed,
\begin{eqnarray}                                
	\tr[X \rho] \geq 0 \hspace{0.15 cm}                             
	\forall \rho \in {\cal T}_{k} &\Rightarrow &
	\tr[X \tr_{\backslash 1}(\pi)] \geq 0 \hspace{0.15 cm} \forall \pi \in 
	{\cal S}_{k} \cap {\cal C}_{1}  \nonumber\\
	&\Rightarrow& \tr[(X \otimes {\id}^{\otimes{( k - 1)}}) \pi] \geq 0 
	\hspace{0.15 cm} \forall \pi \in {\cal S}_{k} \cap {\cal C}_{1}.
	\nonumber\\
\end{eqnarray}                                                
Hence, any element of the 
dual cone of ${\cal T}_k$ can be written as a sum of an
element of the dual cone of ${\cal S}_k$ and an element
of the dual cone of ${\cal C}_1$, which is nothing but
$\Lambda(\id/2 - \phi_2)$. 
\proofend             

The next Lemma is the key result for the proof of the Theorem 2. It makes a connection between separability and the structure of the dual sets $({\cal T}_k)^*$. Before we turn to its formulation 
and proof, let us introduce some notation, departing from 
earlier conventions. This will make render the argument
more transparent, however. In this Lemma, we will
set ${\cal H} := \cc^{2d} \otimes \cc^{2d}$. 
If we have a tensor
product between a $d\times d$-system and a 
$2\times 2$ system, the latter is thought to be
embedded in a $d\times d$-dimensional system.
We denote with $\id$ the identity operator acting on ${\cal H}$. 
The identity operator acting on $\cc^m \otimes \cc^m$, 
for every other $m$ different from $2d$ will be denoted by 
$\id_{m^2}$.

\begin{lemma}[Dual cone of ${\cal T}_k$ and separability] 
\label{crucial}
Let $\sigma \in {\cal D}(\cc^d \otimes \cc^d)$ and 
$k \in \mathbb{N}^*$. If 
\begin{equation}
	\sigma \otimes (\id_{4}/2 - \phi_2) \in ({\cal T}_k)^{*},
\end{equation}	
then $\sigma$ is separable. 
\end{lemma}
\textit{Proof:} 
By Lemma \ref{dual} we can write
\begin{equation}
\sigma \otimes (\id_{4}/2 - \phi_2) \otimes \id^{\otimes{( k - 1)}} = \Lambda(\id_{4}/2 - \phi_2) + Q,
\end{equation}
for some SLOCC map $\Lambda$ and an operator $Q \in {\cal S}_k^*$. Applying the symmetrizing 
operator $\hat{S}$ to both sides of the previous equation, 
we find
\begin{eqnarray}
	\sigma \otimes (\id_{4}/2 - \phi_2) \otimes \id^{\otimes{( k - 1)}} 
	&+& \sum_{j=1}^k \id^{\otimes j} \otimes 
	(\sigma \otimes (\id_{4}/2 - \phi_2))  \nonumber\\
	&\otimes& \id^{\otimes (k - j - 1)}\\
	&=& (\hat{S} \circ \Lambda)(\id_4/2 - \phi_2) 
	+ \hat{S}(Q). \nonumber
\end{eqnarray}
We now multiply both sides from the left 
with $\id \otimes (\id_{d^2} \otimes \ket{0, 0}\bra{0, 0})^{\otimes{( k - 1)}}$ and take the 
partial trace with respect to all systems except the
first $\cc^{2d}\otimes \cc^{2d}$-dimensional subsystem.
Defining
\begin{eqnarray}
	P &:= &\tr_{\backslash 1}[\id \otimes 
	(\id_{d^2} \otimes \ket{0, 0}
	\bra{0, 0})^{\otimes{( k - 1)}} \hat{S}(Q)], \\
	\Upsilon(.) &:=& 
	\tr_{\backslash 1}
	[\id \otimes (\id_{d^2} \otimes 
	\ket{0, 0}\bra{0, 0})^{\otimes{( k - 1)}} 
	(\hat{S}\circ \Lambda)(.)], 
\end{eqnarray}
it follows that
\begin{equation} \label{mas}
	\sigma \otimes (\id_{4}/2 - \phi_2) = 
	\Upsilon(\id_{4}/2 - \phi_2) + P,
\end{equation}
since 
\begin{equation}
	\tr[(\id_{d^2} \otimes \ket{0, 0}\bra{0, 0})(\id_{4}/2 - \phi_2) ]=0.
\end{equation}	
By Lemma \ref{S(Q)=0}, we find that $P \geq 0$. 

The quantum operation $\Upsilon$ can easily be seen to be a SLOCC, 
as it is a concatenation of the SLOCC map 
$\Omega$ with the symmetrizing operation -- which is LOCC -- and finally with the projection of the qubit part of the final $k - 1$ copies in the local state $\ket{0, 0}$,  followed by tracing over them.
Each of these steps can be done locally. 
The statement of the
Lemma then follows from the results presented in 
Ref.\ \cite{Masanes}, where it was shown that Eq.\ 
(\ref{mas}) implies the separability of the bi-partite state
$\sigma\in{\cal D}(\cc^d\otimes \cc^d)$. 
\proofend

\textit{Proof of Theorem 2:} 
Theorem 2 can now be easily established by Lemma \ref{crucial}, together with the argument presented in 
Ref.\ \cite{Masanes}. Let us consider states $\sigma\in
{\cal D}(({\cal H}_{A_{2}}\otimes {\cal H}_{A_{3}})
\otimes ({\cal H}_{B_{2}}\otimes {\cal H}_{B_{3}}))$,
where
\begin{equation}
	{\cal H}_{A_{2}} = {\cal H}_{B_{2}} = 
	\cc^d,\,\,\,
	{\cal H}_{A_{3}} = {\cal H}_{B_{3}} = \cc^2.
\end{equation}
The Hilbert spaces ${\cal H}_{A_1}=\cc^d$ and
${\cal H}_{B_1}=\cc^d$ will serve as the Hilbert spaces
on which the activator $\rho$ is defined.
This might seem like an undesirable complication of 
notation; as in Ref.\ \cite{Masanes}, the discussion of 
the process will become more transparent 
in this way, however.

We aim at activating entanglement.
It hence
suffices to show that for all $k \in \mathbb{N}^*$ there exists 
a $\rho \in {\cal T}_{k}\subset {\cal D}(\cc^d\otimes \cc^d)$ 
and a  SLOCC operation $\Lambda$ such that 
\begin{equation}\label{Hold}
	\tr[\Lambda(\sigma \otimes \rho)\phi_2] > 	
	\tr[\Lambda(\sigma \otimes \rho)]/2. 
\end{equation}
The state $\rho$ then serves as an activator in this
single-copy distillation process.

We are free to show that Eq.\ (\ref{Hold}) is true 
for a particular
choice of a SLOCC operation. This does not necessarily have to
be one that would give the optimal rate, or in the single-copy
regime the optimal overlap, as long as we can show that
the activation has been successful.
We choose $\Lambda$ as follows: 
As a first step, the parties perform a local measurement -- on subsystems $A_1A_2$, $B_1B_2$ -- in a basis of maximally 
entangled states, post-selecting when both 
systems are projected onto the projectors associated with 
the unnormalized state vectors
$|\phi_{A_{1}A_{2}}\rangle = \sum_{i=1}^{d} \ket{i,i}$ and $|\phi_{B_{1}B_{2}}\rangle = \sum_{i=1}^{d} \ket{i,i}$, respectively. The implemented SLOCC is then given by 
\begin{equation}  
        \rho \otimes \sigma 
        \mapsto (A \otimes B)
        (\rho \otimes \sigma ) (A \otimes B)^{\cal y},
\end{equation}
where
\begin{equation}
        A = \bra{\phi_{A_{1}A_{2}}} \otimes \id_{A_{3}}, \hspace{0.3 cm} B = 
        \bra{\phi_{B_{1}B_{2}}} \otimes \id_{B_{3}}, \hspace{0.3 cm} 
\end{equation}
and $|\phi_{A_{1}A_{2}}\rangle$ 
is the state vector of a
maximally entangled state in the Schmidt basis.  

This construction is nothing but the extended Jamiolkowski isomorphism between bi-partite states and non-local 
operations, see, e.g., 
Ref.\ \cite{Ciracjamil}: By performing 
two joint measurements locally on the states 
$\rho$ and $\sigma$, a non-local quantum operation, determined by $\rho$, will be performed in $\sigma$. For our purposes, it is sufficient to consider the following relation:
\begin{equation}\label{Jam}
        \tr[(A \otimes B) (\rho \otimes \sigma)(A \otimes B)^{\cal y}
        Z] = c
        \tr[\rho (\sigma^{T} \otimes Z)],
\end{equation}
for every positive operator $Z$ on
${\cal H}_{A_3}\otimes {\cal H}_{B_3}$ and for some $c>0$.

The condition $\tr[\Lambda(\rho \otimes \sigma)\phi_{2}] > \tr[\Lambda(\sigma \otimes \rho)]/2$ can clearly be written as 
\begin{equation}
        \tr[\Lambda(\rho \otimes \sigma)(\id / 2 - \phi_2)] < 0.
\end{equation}
Hence, from Eq.\ (\ref{Jam}) we get
\begin{equation} \label{sat}
	\tr[ \rho \sigma^T \otimes (\id/2 - \phi_2)] < 0.
\end{equation}
To complete the proof it sufficies to note that by Lemma \ref{crucial}, if $\sigma$ is entangled then there must be a state $\rho \in {\cal T}_k$ satisfying Eq.\ (\ref{sat}). Indeed, if this were not true, then $\sigma^{T}\otimes (\id/2 - \phi_2)$ would have to belong to the dual cone of ${\cal T}_k$, which was shown in Lemma \ref{crucial} to imply the separability of $\sigma$. This proves
the validity of Theorem 2.\proofend

\section{Proof of Corollary 3}

{\it Proof of Corollary 3:}
It is easy to see that it is sufficient to prove that
\begin{equation}
\text{cl}\left(\text{cone}\left(\bigcup_{k \in \mathbb{N}^*} {\cal T}_{k}^{*}\right)\right) = {\cal T}^{*},
\end{equation}
where $\text{cl}(A)$ is the closure of $A$.
Let us first show that 
\begin{equation} \label{eq1a}
\text{cl}\left(\text{cone}\left(\bigcup_{k \in \mathbb{N}^*} {\cal T}_{k}^{*}\right)\right) \subseteq {\cal T}^{*}.
\end{equation}
Choose an element $Y$ of ${\cal T}_k^{*}$. Then $\tr[XY] \geq 0$ for all $X \in {\cal T}_{k}$ and, therefore, for all $X \in \bigcap_{k \in \mathbb{N}^*}{\cal T}_{k} = {\cal T}$. Hence $Y$ is an element of ${\cal T}^{*}$ as well. Thus, for all $k \geq 1$,
\begin{equation}
{\cal T}_k^{*} \subseteq {\cal T}^{*},
\end{equation}
from which follows that
\begin{equation}
\bigcup_{k \in \mathbb{N}^*}{\cal T}_k^{*} \subseteq {\cal T}^{*}.
\end{equation}
As ${\cal T}^{*}$ is a closed convex cone, we get Eq.\ (\ref{eq1a}).

To prove the converse inclusion, we show the following relation
\begin{equation} \label{eq2a}
\left( \bigcup_{k \in \mathbb{N}^*}{\cal T}_k^{*} \right)^{*} \subseteq \bigcap_{k \in \mathbb{N}^*} \text{cone}({\cal T}_{k}).
\end{equation}
Then, using that $\text{cl}(\text{cone}(B)) \subseteq \text{cl}(\text{cone}(A))$ if $A^* \subseteq B^*$ together with the easily established relation
\begin{equation}
\text{cone} \left( \bigcap_{k \in \mathbb{N}^*} {\cal T}_{k}  \right) = \bigcap_{k \in \mathbb{N}^*} \text{cone}({\cal T}_{k}),
\end{equation}
we find the announced result.

Let us then turn to prove Eq.\ (\ref{eq2a}). 
Choose an element $X$ of 
$\left( \bigcup_{k \in \mathbb{N}^*}{\cal T}_k^{*} \right)^{*}$. 
Then 
\begin{equation}
	\tr[XY] \geq 0
\end{equation}
for all $Y \in \bigcup_{k \in \mathbb{N}^*}{\cal T}_k^{*}$, 
which implies that $\tr[XY] \geq 0$ for all $Y \in {\cal T}_k^{*}$
and for all $k \geq 1$. 
Therefore, 
\begin{equation}
	X \in ({\cal T}_k^{*})^{*}, 
\end{equation}
which is equal to $\text{cone}({\cal T}_k)$. 
As this is true for all $k \geq 1$, we arrive at 
Eq.\ (\ref{eq2a}). \proofend

\section{On the existence of NPPT bound entanglement}

Before we conclude this work, 
we would like to comment on the applicability on this 
approach to the conjecture on the existence of bound 
entangled states with a non-positive partial transpose, and in 
particular to Conjecture 1. 
The kind of statement that we would
need is very similar to the one established here: We have
introduced an idea of how to grasp asymptotic entanglement
manipulation in the form of a single-copy activation argument. 
It is clear that if we could prove the validity of 
Lemma \ref{crucial} for the full set ${\cal T}$, then Conjecture 1 
would in fact be true. Indeed, if the activation procedure outlined in the proof of Theorem 2 works for a convex combination of undistillable states, then it has to work at least for one of the states appearing in the convex combination, as it is made explicit by the linearity of Eq.\ (\ref{sat}).

However, although the presented methods seem applicable
to this question, a significant further step seems to be necessary,
and a naive extension of  Lemma \ref{crucial} to 
${\cal T}$ does not seem to work. Indeed, if we 
assume that $\sigma \otimes (\id_4/2 - \phi_2) \in {\cal T}^*$, 
then, by Corollary 3, for every $\varepsilon > 0$, 
there exists 
an integer $n_{\varepsilon}$ such that
\begin{equation}
\sigma \otimes (\id_4/2 - \phi_2) + \varepsilon \id \in ({\cal T}_{k_{\varepsilon}})^*.
\end{equation}

If we followed the steps taken in the proof of Lemma \ref{crucial}, we would find, instead of Eq.\ (\ref{mas}), the following:
\begin{equation}
\sigma \otimes (\id_4/2 - \phi_2) + (n_{\varepsilon} - 1) \varepsilon \id = \Omega_{\varepsilon}(\id/2 - \phi_2) + P_{\varepsilon},
\end{equation}
where $P_{\varepsilon} \geq 0$ and $\Omega_{\varepsilon} $ 
is a SLOCC operation
for every $\varepsilon > 0$. Hence, in order to be able to carry over with the approach similar to one outlined in 
Ref.\ \cite{Masanes}, we would have to be able to show 
that we can choose the sequence 
$\{ n_{\varepsilon} \}$ to be such that
\begin{equation} 
\lim_{\varepsilon \rightarrow 0} (n_{\varepsilon} - 1) \varepsilon = 0.
\end{equation}
Although it could well be the case that such relation hold, we could not find a way either to prove it nor to disprove it, despite considerable
effort.

From a different perspective, it seems that the rate of convergence of an arbitrary element of ${\cal T}^*$ by elements of the inner approximations given by ${\cal T}_k^*$ matters when it comes to the activation properties of the elements of ${\cal T}$. Note that it is exactly the closure in
\begin{equation}\label{CH}
{\cal T}^* = \overline{\bigcup_{k \in \mathbb{N}^*} {\cal T}_k^*}
\end{equation}
the responsible for this behavior. Indeed, Lemma \ref{crucial} can straightforwardly be applied if we require only that
\begin{equation}
\sigma \otimes (\id_4/2 - \phi_2) \in \bigcup_{k \in \mathbb{N}^*} {\cal T}_k^*.
\end{equation}
So the question of the existence of NPPT bound entanglement
can be related and reduced to the question of the
necessity of the convex hull in 
Eq.\ (\ref{CH}). 
   
\section{Summary and conclusions}

In this work, we have introduced the notion of copy-correlated
entanglement distillation. In this setting, one allows for 
correlations between different specimens in entanglement
distillation. We have proven a relationship between
copy-correlated undistillable states and undistillable states,
hence establishing a new way of characterizing the set
of undistillable states. We have also introduced
a new entanglement 
activation result which on one hand generalizes previous ones and 
on the other hand might be of use to the study of the properties 
of the undistillable state set.  

After all, it is not a too unrealistic hope that the methods
this work has introduced may pave an avenue to prove 
the validity of the 
conjecture on the existence of NPPT bound 
entanglement. With new results on 
almost i.i.d. properties of many subsystems of
permutation invariant being just available \cite{RennerNew},
this goal may be within reach. Beyond this specific
question of entanglement distillation, we 
hope that the presented 
methods and tools open up a new 
way of grasping asymptotic entanglement 
manipulation. 

\section{Acknowledgements}
  
We acknowledge fruitful and interesting conversations
with a number of people on these and very closely related
topics, among them 
D.\ Gottesman,
D.\ Gross,
M.\ Horodecki,
Ll.\ Masanes,
B.\ Terhal,
J.\ Oppenheim, 
M.\ Piani,
M.B.\ Plenio,
S.\ Virmani,
K.G.H.\ Vollbrecht, and
R.F.\ Werner. This work has been supported by the
EU (QAP), the Royal Society, 
the QIP-IRC, Microsoft Research, the Brazilian agency CNPq, 
and the EURYI Award Scheme.


\begin{thebibliography}{0}

\bibitem{Werner}
	R.F.\ Werner, Phys.\ Rev.\ A {\bf  40}, 4277  (1989).
	
\bibitem{HorodeckiRMP}
	R.\ Horodecki, P.\ Horodecki, M.\ Horodecki, and
	K.\ Horodecki, quant-ph/0702225.
	
\bibitem{ben} 
	C.H.\ Bennett, G.\ Brassard, S.\ Popescu, B.\ Schumacher, 
	J.A.\ Smolin, 
	and W.K.\ Wootters, Phys.\ Rev.\ Lett.\ \textbf{76}, 722 (1996).

\bibitem{IID}
	The case of arbitrary non i.i.d.\ sequences of states has been 	
	considered, e.g., in Refs.\ \cite{Bowen, Matsumoto}.

\bibitem{Bowen}
 	G.\ Bowen and N.\ Datta, quant-ph/0610199.  

\bibitem{Matsumoto}
 	K.\ Matsumoto, arXiv:0708.3129. 

\bibitem{hor2} 
	M.\ Horodecki, P.\ Horodecki, and R.\ Horodecki, Phys.\ Rev.\ 	
	  Lett.\ \textbf{80}, 5239 (1998);
	Phys.\ Rev.\ A \textbf{60}, 1888 (1999).

\bibitem{Peres}
	A.\ Peres, Phys.\ Rev.\ Lett.\ {\bf 77}, 1413 (1996).
		
			
		 	
\bibitem{ShorNPPT}
		D.P.\ DiVincenzo, P.W.\ Shor, J.A.\ Smolin, B.M.\ Terhal, and 
		A.V.\ Thapliyal, Phys.\ Rev.\ A {\bf 61}, 062312 (2000).

\bibitem{CiracNPPT}
	D.\ D\"ur, J.I.\ Cirac, M.\ Lewenstein, and D.\ Bruss,
	Phys.\ Rev.\ A {\bf 61}, 062313 (2000).
		
\bibitem{ShorCont}
	P.W.\ Shor, J.A.\ Smolin, and B.M.\ Terhal, 
	Phys.\ Rev.\ Lett.\ {\bf 86}, 2681  (2001).

\bibitem{NPPTAct} 
  	T.\ Eggeling, K.G.H.\ Vollbrecht, R.F.\ Werner, 
	and M.M.\ Wolf, 
  	Phys.\ Rev.\ Lett.\ \textbf{87}, 257902 (2001).  

\bibitem{NPPTAct2}		 		
	K.G.H.\ Vollbrecht and M.M.\ Wolf,
	Phys.\ Rev.\ Lett.\ {\bf 88}, 247901 (2002).

\bibitem{Masanes} 
	Ll.\ Masanes, 
	Phys.\ Rev.\ Lett.\ \textbf{96}, 150501 (2006).

\bibitem{hor3} 
	P.\ Horodecki, M.\ Horodecki, and R.\ Horodecki, 
	Phys.\ Rev.\ Lett.\ \textbf{82}, 1056 (1999).
	
\bibitem{Activation}
 	B.\ Kraus, M.\ Lewenstein, and J.I.\ Cirac,
	Phys.\ Rev.\ A {\bf 65}, 042327 (2002).

\bibitem{Ishizaka1}
 	S.\ Ishizaka, Phys.\ Rev.\ Lett.\ \textbf{93}, 190501 (2004). 

\bibitem{Ishizaka2}
  	S.\ Ishizaka and M.B.\ Plenio, 
	Phys.\ Rev.\ A \textbf{71}, 052303 (2005). 
		
\bibitem{Masanes2}
	Ll.\ Masanes,
	quant-ph/0510188.

\bibitem{Brandao1}
	F.G.S.L. Brand\~ao, Phys.\ Rev.\ A {\bf 76}, 030301 (2007).
	
\bibitem{Old}
	J.\ Eisert, T.\ Felbinger, 
	P.\ Papadopoulos, M.B.\ Plenio, and M.\ Wilkens,
	Phys.\ Rev.\ Lett.\ {\bf 84}, 1611 (2000).

\bibitem{Datta}
	G.\ Bowen and N.\ Datta, quant-ph/0610199.
	
\bibitem{Convex}
	R.T.\ Rockafellar, {\it Convex analysis} (Princeton 
	University 		
	Press, Princeton, 1970).

\bibitem{Comm}
	The set of single-copy undistillable states forms a convex set
	and membership can easily be tested in a one-sided test
	with a witness, see, e.g., Ref.\ \cite{Clarisse}. Also, ideas of
	relaxations \cite{Do,Je,Br} to robust semi-definite programs can 	be applied to 
	study the distillability of Werner states \cite{Vianna}. 

\bibitem{Simon}
	R.\ Simon, quant-ph/0608250.
	
 \bibitem{Clarisse}
 	L.\ Clarisse,
	Quant.\ Inf.\ Comp.\, {\bf 6}, 539  (2006).
	
\bibitem{Do}
	A.C.\ Doherty, P.A.\ Parrilo, and F.M.\ Spedalieri,
	Phys.\ Rev.\ A {\bf 69}, 022308 (2004).

\bibitem{Je} 
	J.\ Eisert, P.\ Hyllus, O.\ G{\"u}hne, and M.\ Curty, 
	Phys.\ Rev.\ A \textbf	{70}, 062317 (2004).

\bibitem{Br}
	F.G.S.L.\ Brand\~ao and R.O.\ Vianna,
	Phys.\ Rev.\ Lett.\ {\bf 93}, 220503 (2004).

\bibitem{Vianna}
	R.O.\ Vianna and A.C.\ Doherty,
	Phys.\ Rev.\ A {\bf 74}, 052306 (2006).
	
%

	


\bibitem{Kuratowski}
	C.\ Kuratowski, Topologie Vol.\ I, PWN Warszawa (1958).

\bibitem{Problem}
	Problem 2, {\it Undistillability implies ppt?}, on the  
	problem page maintained by R.F.\ Werner's quantum
	information group at the Technical
	University of
	Braunschweig, 
	http://www.imaph.tu-bs.de/qi/problems/2.html.	
	
\bibitem{Fuchs}
  	C.A.\ Fuchs, R.\ Schack, and P.F.\ Scudo, 
  	Phys.\ Rev.\ A \textbf{69}, 062305 (2004).

\bibitem{Christandl}
 	M.\ Christandl, R.\ K\"onig, G.\ Mitchison, and R.\ Renner,
 	quant-ph/0602130.

\bibitem{Renner}
  	R.\ K{\"o}nig and R.\ Renner,
  	J.\ Math.\ Phys.\ \textbf{46}, 122108 (2005).	   

\bibitem{Tomas}
	T.M. Cover and J.A. Thomas, 
	\textit{Elements of information theory}
	(John Wiley and Sons, New York, 1991).

\bibitem{lasserre}
  	O.\ Hernandez-Lerma and J.B.\ Lasserre, 
  	J.\ Conv.\ Anal.\ \textbf{4}, 164 (1997). 
		
\bibitem{Ciracjamil}
  	J.I.\ Cirac, W.\ D\"ur, B.\ Kraus, and M.\ Lewenstein,
  	Phys.\ Rev.\ Lett.\ {\bf 86}, 544 (2001).	


		     	
  

\bibitem{RennerNew}
	R.\ Renner, quant-ph/0703069.

\end{thebibliography}
\end{document}